\documentstyle[aps,preprint,epsfig,floats]{revtex}
\tightenlines

\begin{document}

\hfuzz=4pt

\preprint{\font\fortssbx=cmssbx10 scaled \magstep2
\hbox to \hsize{
\hbox{\fortssbx University of Wisconsin - Madison}
\hfill$\vcenter{\tighten
                \hbox{\bf MADPH-98-1045}
                \hbox{\bf hep-ph/9802421}
                \hbox{February 1998}}$}}

\title{\vspace*{.25in}  
Inelastic Photoproduction at HERA:\\
 a Second Charmonium Crisis?}

\author{O.\ J.\ P.\ \'Eboli$^{1,2}$,
E.\ M.\ Gregores$^1$, and F.\ Halzen$^1$}
\address{\vspace*{3mm}
$^1$Department of Physics, University of Wisconsin \\
Madison, WI 53706, USA\\[3mm]
$^2$Instituto de F\'{\i}sica Te\'orica,
Universidade Estadual Paulista\\
Rua Pamplona 145, S\~ao Paulo, SP 01405-900, Brazil}

\maketitle
\thispagestyle{empty}

\begin{abstract}
\vskip-5ex 

The measurement of the inelastic photoproduction of charmonium at HERA seems
to have ignited a new charmonium crisis. The (already discredited) color
singlet model fits the data for large charmonium energy fraction $z$, where
the NRQCD model qualitatively fails. We here point out that by the
straightforward inclusion of color singlet and octet processes in the soft
color (color evaporation) scheme, the HERA data can be accommodated for all
$z$.  We anticipate that the color singlet model will fail at low $z$, as it
does in hadroproduction.

\end{abstract}
\draft

\newpage


\section{Introduction}

The once conventional treatment of color in perturbative QCD calculations,
i.e., the color singlet model (CSM), has run into a serious impasse in
describing the data on the production of charmonium states at the Fermilab
Tevatron \cite{com-tev}. Specific proposals to solve this problem agree on the
basic elements: i) onium production is a two-step process where a heavy quark
pair is produced first, followed by the nonperturbative formation of the
asymptotic states, and ii) color octet as well as singlet $c\bar{c}$ states
contribute to the production of charmonia, as clearly demonstrated by the data
\cite{com-tev,us-2}.  Neglecting octet contributions is the source of the
failure of the CSM. Two formalisms have been proposed to remedy this
mistake. The basic assumption of the soft color approach \cite{cem,us-1} is
that no observable dynamics is associated with the soft processes that connect
the color of the perturbative charm pair with the colorless charmonium bound
state. This scheme, although far more restrictive than other proposals,
successfully accommodates all features of charmonium and bottomonium production.  The original QCD computations\cite{cem} referred to it as the color evaporation model (CEM), and identified it as the most logical way to incorporate color constraints into perturbative QCD calculations. The data suggests it still is. It correctly predicts the energy and final state momentum dependence of charmonium and bottomonium
hadro- and photoproduction at all energies, as well as their production in
electron-positron colliders. It makes the sweeping prediction that $\psi$'s
are produced unpolarized, a signature of the absence of dynamical final state
effects. If such effects were discovered, there is a possible scheme,
successful in the theoretical treatment of decays, to accommodate them: the
non-relativistic QCD (NRQCD or COM, for color octet model)
\cite{bbl}. Applying NRQCD to the bulk of the data has been a challenge and
has, in fact, run into difficulties with recent HERA data.

The measurement \cite{hera} of the inelastic photoproduction of charmonium at
high energy has triggered a new charmonium crisis. The CSM, which
qualitatively fails to describe the hadroproduction of charmonium, fits the
HERA data for large charmonium energy fraction $z$ \cite{kramer-csm}, where
the NRQCD model qualitatively fails \cite{kramer-com}.  In this work we show
that the CEM accommodates the data at all $z$. Moreover, we anticipate that
the CSM will fail at low $z$, as it does in hadroproduction, since the
$J/\psi$ photoproduction is mediated by resolved photons in that region.

The evidence is compelling that Nature operates according to the color
evaporation scheme. The CEM formalism simply predicts that, up to color and
normalization factors, the energy, $x_F$- and $p_T$-dependence of the
cross section are identical for the production of onium states and $D
\bar D$ pairs. This is indeed the case \cite{us-2,us-1,vogt}. 
The sum of the cross sections of all onium and open charm states is described
by \cite{us-1}
\begin{equation}
\sigma_{\rm onium} = \frac{1}{9} \int_{2 m_c}^{2 m_D} dM_{c \bar{c}}~
\frac{d \sigma_{c \bar{c}}}{dM_{c \bar{c}}} \; ,
\label{sig:on}
\end{equation}
and
\begin{eqnarray}
\sigma_{\rm open} &=& \frac{8}{9}  \int_{2 m_c}^{2 m_D} dM_{c \bar{c}}~
\frac{d \sigma_{c \bar{c}}}{d M_{c \bar{c}}}
+ \int_{2 m_D} d M_{c \bar{c}}~\frac{d \sigma_{c \bar{c}}}{dM_{c \bar{c}}} 
\; ,
\label{sig:op}
\end{eqnarray}
where the cross section for producing heavy quarks, $\sigma_{c
\bar{c}}$, is computed perturbatively, irrespective of the color of
the $c \bar{c}$ pair. $M_{c \bar{c}}$ is the invariant mass of the $c \bar{c}$
pair.  The CEM assumes a factorization of the production of the $c\bar{c}$
pair, which is perturbative and process-dependent, and the materialization of
this pair into a charmonium state by a mechanism that is nonperturbative and
process independent.  This assumption is reasonable, given that the
characteristic time scales of the two processes are very different: the time
scale for the production of the pair is the inverse of the heavy-quark mass,
while the formation of the bound state is associated to the time scale
$1/\Lambda_{\rm QCD}$. This approach to color is also suggestive of the
unorthodox prescription for the production of rapidity gaps in deep inelastic
scattering and in the production of hard jets at the Tevatron
\cite{bh,faro,D0gap}.

Comparison with the $\psi$ data requires
knowledge of the fraction $\rho_\psi$ of produced onium states that
materialize as $\psi$'s, {\em i.e.,}
\begin{equation}
\sigma_\psi = \rho_\psi~\sigma_{\rm onium} \; ,
\label{frac}
\end{equation}
where $\rho_\psi$ is assumed to be a constant.  This assumption is in
agreement with the low-energy data \cite{gksssv,schuler}. Notice that a single
nonperturbative factor $\rho$ describes a given charmonium state, regardless
of the spin and orbital angular momentum of the charm pair. Analysis of
charmonium photoproduction allowed us to determined that $\rho_\psi \simeq
0.43$--$0.5$ \cite{us-2}. Even this parameter can be accounted for on the
basis of statistical counting of final states \cite{ingelman}, as expected in
a scheme with no final state dynamics.

Quantitative tests of color evaporation are made possible by the fact that all
$\psi$-production data, i.e.\ photo-, hadroproduction, $Z$-decay, etc., are
described in terms of the single parameter $\rho_\psi$ describing the
frequency by which a charm pair turns into a $J/\psi$ via the final state
color fluctuations. Its value should, in the spirit of the model, be the same
for all processes. We have demonstrated \cite{us-2} the quantitative precision
of the color evaporation scheme by showing how it accommodates all
measurements, including the high energy Tevatron and HERA data, which have
represented a considerable challenge for other computational schemes. Its
parameter-free prediction of the rate for $Z$-boson decay into $\psi$'s is an
order of magnitude larger than the color singlet model and consistent with
data \cite{us-z0}.


\section{Results and Conclusions}

Application of the the soft color scheme to inelastic charmonium  
photoproduction is straightforward. The photoproduction cross section is  
related to the $ep$ cross section by
\begin{equation}
 \sigma_{e p\rightarrow J/\psi X}(s)=
\int_{y_{min}}^{y_{max}} \int_{Q^2_{min}}^{Q^2_{max}}
\Phi(y,Q^2)~ \sigma_{\gamma p\rightarrow J/\psi X}(W)\,dy\,dQ^2 \; ,
\end{equation}
where $\sigma_{\gamma p\rightarrow J/\psi X}(W)$ is a function of the
center--of--mass energy of the $\gamma p$ system, and $\Phi$ is the
distribution function of photons with virtuality $Q^2$.  The $Q^2$ integration
ranges from $Q^2_{min}=M_e^2y^2/(1-y)$ to $Q^2_{max}$ ( = 4 GeV$^2$). The
limits for the $y$ integral are $y_{max/min}= W^2_{max/min} / s$, where
$W_{max/min}$ represent the maximum/minimum experimental values of $W$ in the
range $40 < W < 140$ GeV \cite{hera}. We used the following photon distribution
function:
\begin{equation}
\Phi(y,Q^2) =
\frac{\alpha}{2\pi yQ^2}\left[1+(1-y)^2-\frac{2M_e^2 y^2}{Q^2}\right]
\; .
\end{equation}

According to the parton model, the cross section for the $J/\psi$
photoproduction at a given center--of--mass energy $W$ is
\begin{equation}
 \sigma_{\gamma p\rightarrow J/\psi X}(W)=\int dx_A\int dx_B~
f_{A/\gamma}(x_A)f_{B/p}(x_B)~\hat\sigma_{AB\rightarrow J/\psi X}(\hat s)
\; ,
\end{equation}
where the subprocess cross section $\hat\sigma$ is given by Eqs.\
(\ref{sig:on}) and (\ref{frac}) in the CEM.  Here, $\sqrt{\hat
s}=\sqrt{x_A x_B} \, W$ is the center--of--mass energy of the
subprocess $AB\rightarrow J/\psi X$. $f_{A/\gamma}$ ($f_{B/p}$) is
the distribution function of the parton $A$ ($B$) in the photon
(proton).  For direct photon interactions $(A=\gamma)$ we have
$f^{A/\gamma}(x_A)=\delta(x_A-1)$.

The fraction of photon energy transferred to the $J/\psi$ in the
proton rest frame is given by the Lorentz invariant:
\begin{equation}
 z \equiv \frac{P_{J/\psi}\cdot P_{p}}{P_{\gamma}\cdot P_{p}} \; ,
\end{equation}
where $P_{J/\psi,\gamma,p}$ is the four-momentum of the $J/\psi$, photon, and
proton, respectively. Following the cuts applied by the experiments
\cite{hera}, we impose  that the transverse momentum
of the $J/\psi$ is larger than 1 GeV.

\begin{figure}
\begin{center}
\mbox{\epsfig{file=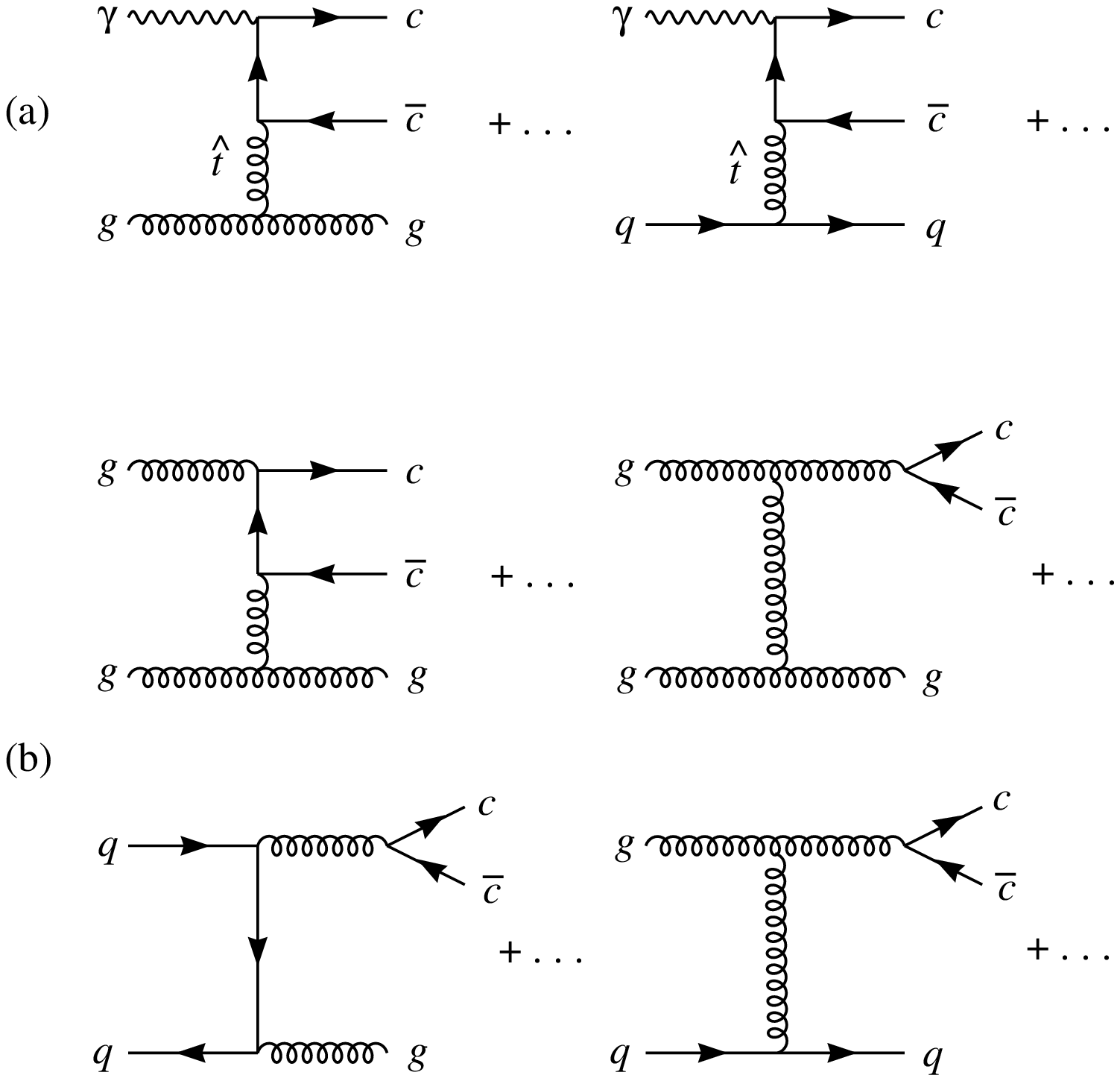,width=0.8\linewidth}}
\end{center}
\caption{${\cal O}(\alpha_{em}\alpha_s^2)$ processes contributing 
to direct (a) and resolved (b)  charmonium photoproduction.}
\label{fig:feyn}
\end{figure}

$\gamma g \rightarrow c \bar{c}$ is the leading order process. It is, however,
only important for large values of $z$. It was, in fact, the process used in
Ref.\ \cite{us-2} to empirically determine the value of $\rho_\psi$. For the
range of $z$-values we are focusing on here, the direct photon contribution
is dominated by
\begin{eqnarray*}
\gamma~ g &\rightarrow& g~ c~ \bar{c} \; ,
\\
\gamma~ q &\rightarrow& q~ c~ \bar{c} \; ,
\end{eqnarray*}
see Fig.~\ref{fig:feyn}. The charm quark pair in $\gamma g$ fusion is produced
in color singlet and octet configurations, while $\gamma q$ fusion leads to
colored pairs only. These processes have to be evaluated with some caution
because for soft values of the gluon transfer momentum $\hat{t}$, see
Fig.~\ref{fig:feyn}, the diagram represents the QCD evolution of the initial
state distribution functions. In order to avoid double-counting one imposes
$|\hat t|<(2m_c)^2$ \cite{flavor}. We also included resolved photon processes,
which proceed via quark--quark, quark--gluon, and gluon--gluon fusion into $c
\bar{c} +$ quark (gluon). The scattering amplitudes were evaluated using the
packages MADGRAPH \cite{madg} and HELAS \cite{helas}.

\begin{figure}
\begin{center}
\mbox{\epsfig{file=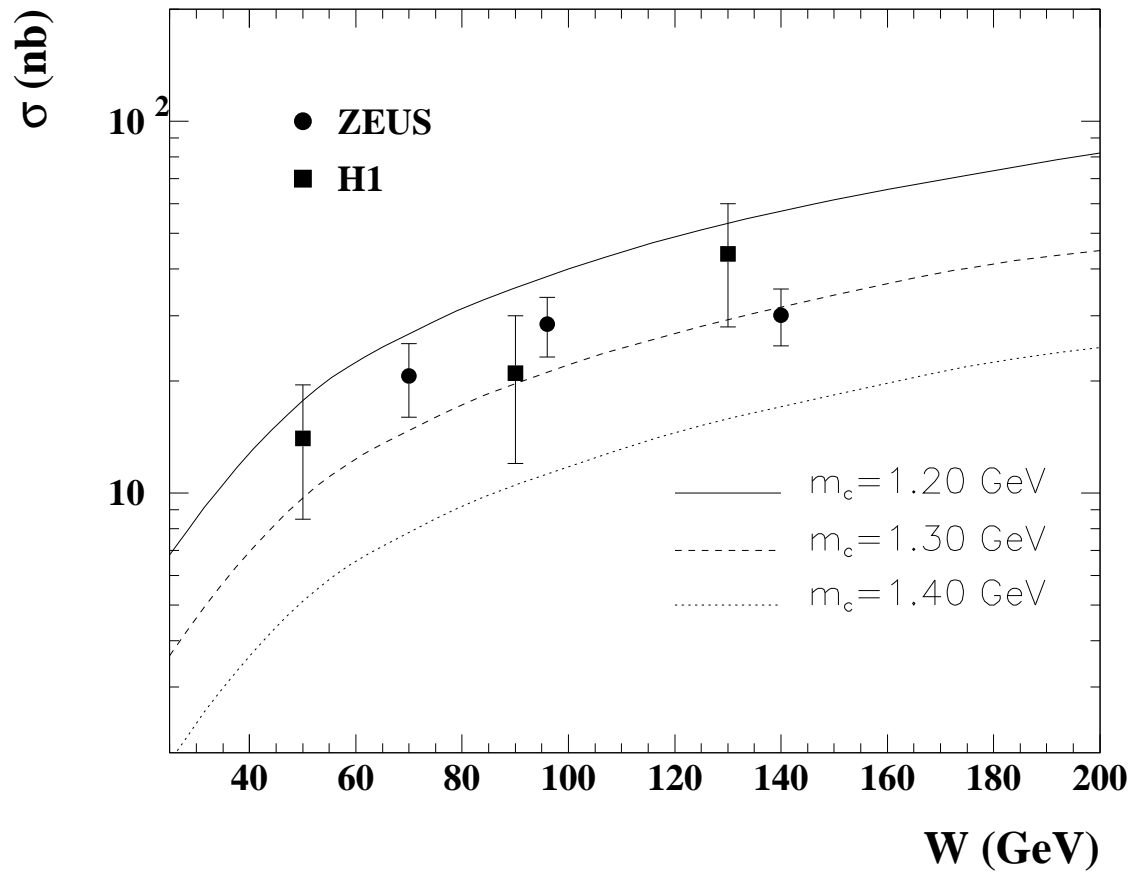,width=0.9\linewidth}}
\end{center}
\caption{Inelastic $J/\psi$ cross section for $p_T>1$ GeV
and $0.4<z<0.8$. We used the GRV94-LO structure function, $\Lambda^{(4)}=300$
MeV, renormalization scale $\mu_R=m_c$, and factorization scale $Q^2=\hat s$.}
\label{fig:xsec}
\end{figure}

The resulting charmonium photoproduction cross section is shown in
Fig.~\ref{fig:xsec} as a function of the center--of--mass energy $W$ for
$0.4<z<0.8$. We fixed $\rho_\psi=$ 0.43, and used GRV distribution functions
for the proton \cite{grv-p} and photon \cite{grv-g}. The adopted
renormalization scale is the charm--quark mass $m_c$ for $\Lambda^{(4)}=300$
MeV, while the factorization scale is $\sqrt{\hat{s}}$. We varied $m_c$
between 1.2 and 1.4 GeV. The relatively large uncertainty in our prediction
indicates the need to include higher order corrections. This is expected
because ${\cal O}(\alpha_{em} \alpha_s^2)$ processes, although formally higher
order, represent the dominant mechanism for $z\ne 1$ charmonium
production. Nevertheless, it is important to notice that it is not a challenge
to describe the data.

\begin{figure}
\begin{center}
\mbox{\epsfig{file=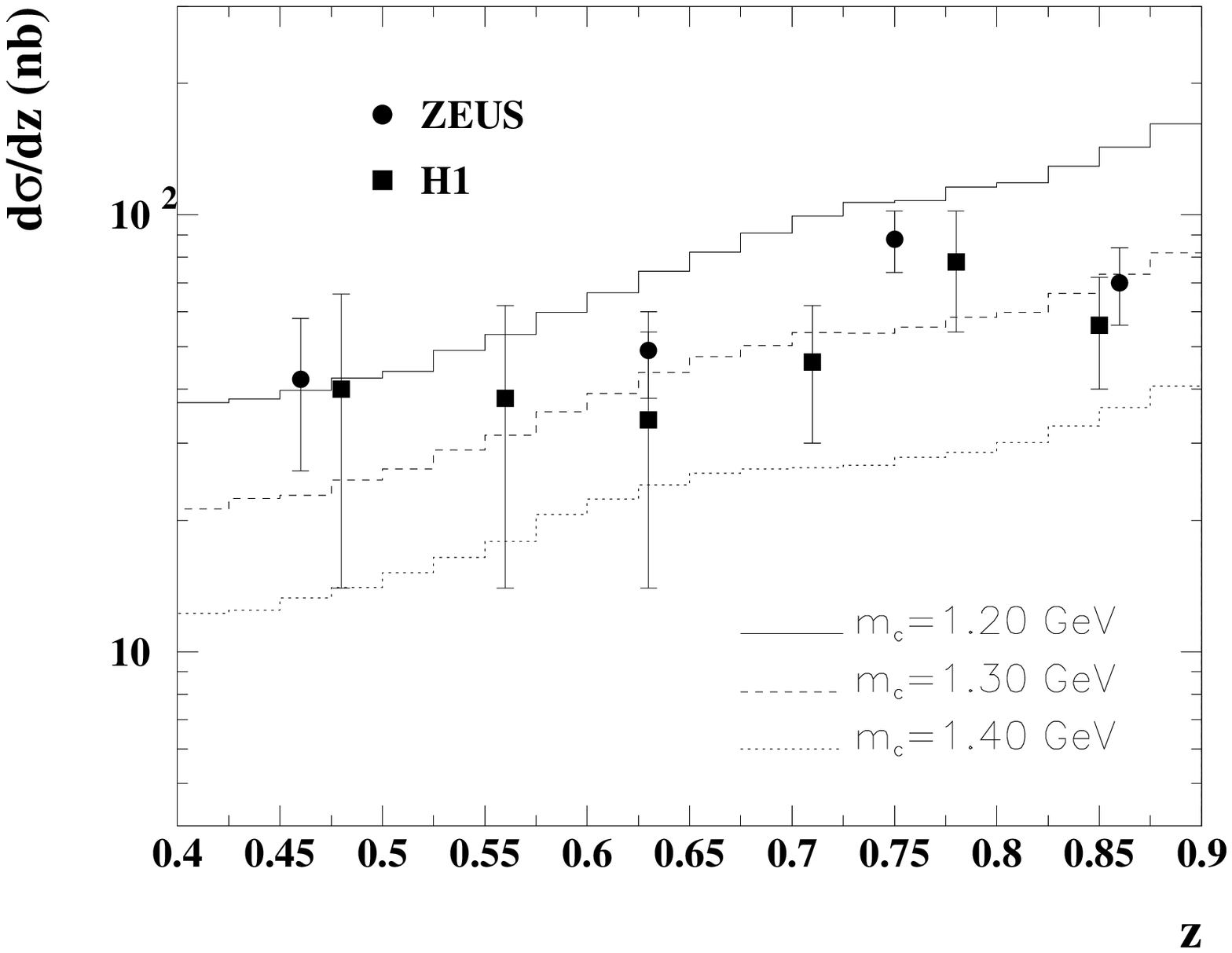,width=0.9\linewidth}}
\end{center}
\caption{Inelastic $J/\psi$ production as function of the
inelasticity parameter $z$ for $p_T>1$ GeV.  We chose the parameters as in
Fig.~\protect\ref{fig:xsec}.}
\label{fig:zdep}
\end{figure}

Having fixed any freedom in the definition of the perturbative series, we
confront the CEM calculation with the $z$--dependence of the cross section for
the inelastic $J/\psi$ photoproduction in Fig.~\ref{fig:zdep}. This figure
shows that CEM successfully passes the test where NRQCD is at variance with
the data\cite{kramer-com}.  The result supports the underlying assumption of CEM that the probability of charmonium formation from perturbative $c\bar{c}$ pairs is independent of the color state, angular momentum and spin of the charm quark pair.

\begin{figure}
\begin{center}
\mbox{\epsfig{file=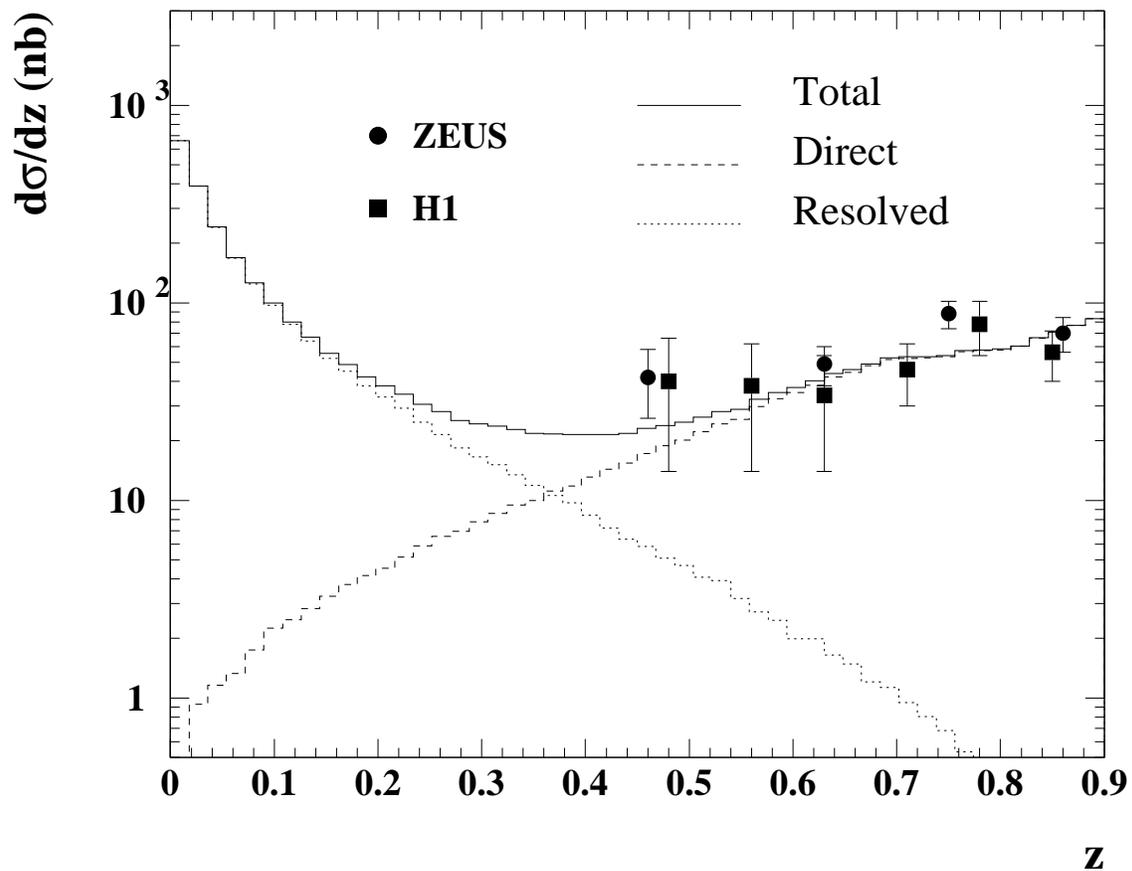,width=0.9\linewidth}}
\end{center}
\caption{Inelastic $J/\psi$ production as function of the
inelasticity parameter $z$ for $p_T>1$ GeV.  We chose the parameters as in
Fig.~\protect\ref{fig:xsec} and fixed $m_c = 1.3$ GeV.}
\label{fig:zres}
\end{figure}

Finally, we show in Fig.~\ref{fig:zres} the direct--photon, resolved--photon,
and total contributions to the $z$--dependence of the inelastic $J/\psi$
photoproduction.  The inclusion of resolved processes strongly increases the
$J/\psi$ yield when $z\rightarrow 0$. The measurement of the $J/\psi$
production in this region can seal, once more, the fate of the CSM, since we
anticipate that the omission of color octet gluon--gluon fusion processes will
result in too small cross sections, as already observed in the Tevatron
data. In principle, this is remedied in the NRQCD framework which, however,
fails for larger values of $z$'s.


\acknowledgments
\unskip\smallskip  
O.~J.~P.~E.~is grateful to the Physics Department of University of Wisconsin,
Madison for its kind hospitality.  This research was supported in part by the
University of Wisconsin Research Committee with funds granted by the Wisconsin
Alumni Research Foundation, by the U.S.\ Department of Energy under grant
DE-FG02-95ER40896, by Funda\c{c}\~{a}o de Amparo \`a Pesquisa do Estado de
S\~ao Paulo (FAPESP), and by Conselho Nacional de Desenvolvimento
Cient\'{\i}fico e Tecnol\'ogico (CNPq).


\newpage


\end{document}